\documentclass{aa}  

\usepackage{graphicx}
\usepackage{hyperref}

\usepackage{txfonts}
\begin{document}

   \title{Exploring the nature of an ultraluminous X-ray source in NGC 628}

   \author{H. Avdan
          \inst{1}
          \and
          S. Avdan\inst{1}
          }

   \institute{Adiyaman University, Department of Physics, 02040 Adiyaman, T\"{u}rkiye\\
              \email{avdan.hsn@gmail.com}
}

   \date{}

 
  \abstract
   {}
   {In this work, we study the X-ray spectral and temporal properties of an ultraluminous X-ray source (ULX) in NGC 628 by using multi-epoch archival X-ray data. The physical parameters were estimated in each epoch in order to constrain the nature of the compact object in the system. Also, the optical counterpart candidates of the ULX were examined using the archival {\it Hubble} Space Telescope (HST)/Wide Field Camera 3 (WFC3) data.} 
   {{\it XMM-Newton}, {\it Chandra}, and {\it Swift} data were used to create the long-term light curve (which covers a period of 22 years) and perform the spectral analysis. Lomb-Scargle periodograms of the source were constructed to examine the short-term variability in each epoch. In order to search for an optical counterpart in the HST/WFC3 images, a relative astrometric correction was initially applied to the {\it Chandra} and HST/WFC3 images.}
   {The X-ray flux of the source changes by a factor of $\sim 200$ throughout the observations. The previously detected quasi-periodic signal (in the range of 0.1$-$0.4 mHz) was confirmed by using the Lomb-Scargle method. After astrometric correction, two optical counterpart candidates were detected for the source. The obtained spectral energy distributions in the optical band for both candidates indicate that the optical emission is dominated by the irradiation of the accretion disc. Considering the best-fit model parameters of the multi-colour disc black-body model, we derived the mass of the black hole in the system as being in the range of (3$-$16) $M_{\odot}$. Nonetheless, the long-term variability and the spectral transitions in the hardness-luminosity diagram make it difficult to rule out the neutron star scenario.}
   {}

   \keywords{ galaxies: individual: NGC 628  --
                X-rays: binaries
               }

   \maketitle
%

\section{Introduction}
Ultraluminous X-ray Sources (ULXs) are generally defined as extra-Galactic, point-like X-ray sources with X-ray luminosities exceeding the Eddington limit for a typical stellar mass black hole (BH) of 10 M$_{\odot}$  ($\sim$2$\times$10$^{39}$ erg s$^{-1}$), and they are located outside their host galaxy nucleus. Even though the true nature of ULXs is still being debated, there are several basic models in the literature that explain the high luminosities of these sources. To explain their high luminosity, ULXs were initially thought to be systems that contain an intermediate-mass black hole (IMBH; 10$^{2}$-10$^{5}$ M$_{\odot}$) accreting at sub-Eddington rates \citep{1999ApJ...519...89C, 2004ApJ...614L.117M, 2009Natur.460...73F}. In contrast, some studies have suggested that these systems contain stellar-mass black holes (SMBHs) and are powered by super-Eddington accretion and/or strong beamed emission \cite[e.g.][]{2001ApJ...552L.109K, 2011NewAR..55..166F, 2013ApJ...773L...9W, 2015MNRAS.447.3243M}. Recently, the detection of pulsation in the M82 X-2 revealed that ULXs may contain neutron stars \citep{2014Natur.514..202B}. With the increasing number of pulsating ULXs (PULXs) that have been detected \citep{2016ApJ...831L..14F, 2017Sci...355..817I, 2017MNRAS.466L..48I, 2017ApJ...836..113P, 2017A&A...608A..47K, 2018ApJ...856..128W, 2019MNRAS.488L..35S, 2020ApJ...895...60R, 2021A&A...649A.104G}, it is thought that most of the ULXs could be super-Eddington accretion systems containing SMBHs or neutron stars rather than IMBHs.

Investigating the X-ray spectral and timing properties of ULXs provides important information for constraining the mass of the compact object (accretor) and understanding the accretion regime. The spectrum of the vast majority of ULXs differs from Galactic black hole binaries (GBHBs), where the accretion usually occurs at a sub-Eddington rate. Typically, ULXs have a soft excess coupled with a spectral curvature below 10 keV. In contrast to ULXs, this curvature is generally seen at higher energies (>10 keV) in GBHBs. Besides, the super-Eddington models predict strong outflows \cite[e.g.][]{2007MNRAS.377.1187P}, and the recent detections of blueshifted lines in the spectra of many ULXs when using the data taken with the Reflection Grating Spectrometers (RGS) instrument aboard \textit{XMM-Newton} have revealed the existence of outflows \cite[e.g.][]{2016Natur.533...64P}. Thus, the spectra of ULXs support the super-Eddington accretion scenario rather than sub-Eddington accretion \citep[and references therein]{2023arXiv230200006P, 2023AN....34420134P}. In addition, ULXs exhibit different spectral states that are generally defined in three categories: soft ultraluminous, hard ultraluminous, and broadened disc \citep{2013MNRAS.435.1758S}. When the spectrum of the source is modelled with a thermal plus a power-law component, the spectral state is classified as soft ultraluminous ($\Gamma$ > 2) or hard ultraluminous ($\Gamma$ < 2), depending on the spectral index. If the spectrum is dominated by the disc component, its spectral state is defined as a broadened disc \citep{2013MNRAS.435.1758S}. Moreover, \cite{2013MNRAS.435.1758S} found that the ULXs in the hard and soft ultraluminous spectral state are generally brighter. 

The optical counterparts of ULXs also play an important role in estimating the spectral class and mass of the companion star and origin of the optical emission itself. Studies using the {\it Hubble} Space Telescope (HST) data show that the optical counterparts of most ULXs are faint, and usually their apparent magnitudes are in the range of 21 $\leq$ m$_{V}$ $\leq$ 26 \citep{2021AstBu..76....6F}. Since their visual magnitudes are very dim and they are usually located in crowded regions (such as star-forming regions), the number of ULXs that were associated with an optical counterpart is only around 30 \citep{2017ARA&A..55..303K, 2021AstBu..76....6F}. Considering the luminosity and colour values of their optical counterparts, the companion stars of many ULXs are massive stars with an O or B spectral type; thus, they can be classified as high-mass X-ray binaries \citep{2002ApJ...580L..31L, 2004AAS...20510403L, 2006ApJ...645..264T, 2013ApJS..206...14G}. On the other hand, some ULXs have companion stars with an A or late type, and they can be classified as low-mass X-ray binaries \citep{2011ApJ...737...81T, 2016MNRAS.455L..91A, 2019ApJ...875...68A}.

The aim of this paper is to study the X-ray properties of the ULX (hereafter X-1) in NGC 628 (M74) in detail. We re-analysed the {\it XMM-Newton} and {\it Chandra} observations that were previously used in the literature to study X-1 as well as the newer {\it XMM-Newton}, {\it Chandra}, and {\it Swift} archival data, which had not been used for this purpose before. Additionally, the optical counterpart of X-1 was searched for using the HST/Wide Field Camera 3 (WFC3) archival data. NGC 628 is a star-forming, face-on spiral galaxy with an inclination angle of $\sim$ (5$-$7)$^{\circ}$ \citep{1984A&A...132...20S} at a distance of 9.7 Mpc \citep{1988Sci...242..310T}. X-1 is located in a spiral arm and is to the south-east of the nucleus. X-1 was uncatalogued in the {\it ROSAT} and {\it Einstein} catalogues, and it was classified for the first time as a bright transient X-ray source with a luminosity of $5\times10^{39}$ erg s$^{-1}$ in the {\it XMM-Newton} observation by \cite{2002ApJ...572L..33S}. Later, \cite{2005ApJ...630..228K} investigated the temporal and spectral variation of X-1 using two {\it Chandra} observations and one from {\it XMM-Newton}. In these observations, the luminosity of X-1 varied from $5 \times 10^{38}$ to $1.2 \times 10^{40}$ erg s$^{-1}$ across about eight months. This extreme variability and spectral state transition of the source are similar to Galactic microquasars, which supports the idea that the origin of the high luminosity of X-1 could be relativistically beamed jets. Also, the authors speculated that the source may contain an IMBH based on the calculated low disc temperature values. \cite{2005ApJ...621L..17L} determined a two-hour quasi period in X-1 by using two {\it Chandra} and two {\it XMM-Newton} observations. Then they calculated the mass of the BH in the system as $\sim$ 10$^{4}$ M$_{\odot}$ using the $f_{b}$-M$_{\medbullet}$ scaling relation, where $f_{b}$ is the break frequency.

This paper is organised as follows. The observations and data reduction steps are described in Sect. \ref{obs-reduct}; the result of the X-ray and optical analyses and the discussions are given in Sect. \ref{results}; and the summary is given in Sect. \ref{summary}.


\section{Observations and data reductions}
\label{obs-reduct}

\begin{figure*}
\begin{center}
\includegraphics[angle=0,scale=0.25]{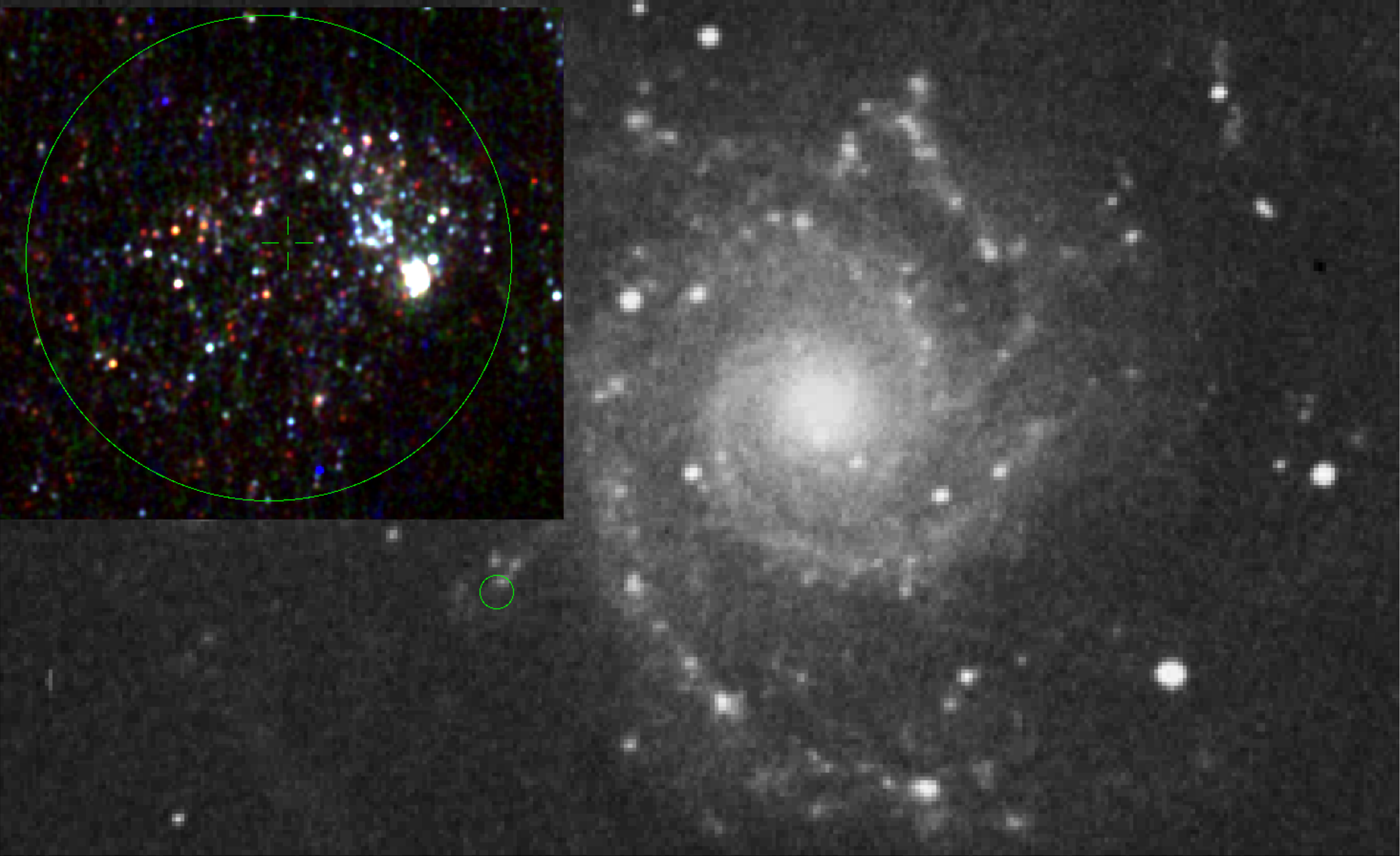}
\caption{Sloan Digital Sky Survey and HST true colour (zoomed) images of NGC 628 (red: F555W; green: F336W; blue: F275W). The green circle represents a 7$\arcsec$ area around X-1, and the corrected position of X-1 is shown with green bars in the zoomed-in HST image.}
\label{F:sdss}
\end{center}
\end{figure*}

NGC 628 was observed multiple times with {\it XMM-Newton}, {\it Chandra}, and {\it Swift} over 22 years. It has also been observed with different HST cameras. We used the F275W, F336W, F555W, and F814W data taken with the WFC3 to investigate the optical counterpart of the ULX X-1. The details of the observations used in this study are summarised in Table \ref{Obs_log}. The Sloan Digital Sky Survey (SDSS) image of NGC 628 and the HST image around X-1 are given in Fig. \ref{F:sdss}.

\begin{table*}
        \centering
        \caption{Archival observations used in this work.}
        \label{tab:obs-log}
        \begin{tabular}{llccc} 
                \hline
        \hline
Observatory & Label & ObsID (Filter) & Date & Exposure  \\
 & & & & (ks) \\
\hline
{\it XMM-Newton} & XM1 & 0154350101 & 2002 Feb 01 & 37  \\
                 & XM2 & 0154350201 & 2003 Jan 07 & 25  \\
                 & XM3 & 0864270101 & 2021 Jan 13 & 112 \\
\hline
{\it Chandra} & C1 & 2057 & 2001 Jun 19 & 46 \\
              & C2 & 2058 & 2001 Oct 19 & 46 \\
              & C3 & 4753 & 2003 Nov 20 & 5 \\
              & C4 & 4754 & 2003 Dec 18 & 5 \\
              & C5 & 14801 & 2013 Aug 21 & 10 \\
              & C6 & 16000 & 2013 Sep 21 & 44 \\
              & C7 & 16001 & 2013 Oct 07 & 15 \\
              & C8 & 16484 & 2013 Oct 10 & 15 \\
              & C9 & 16485 & 2013 Oct 11 & 9 \\
              & C10 & 16002 & 2013 Nov 14 & 38 \\
              & C11 & 16003 & 2013 Dec 15 & 40 \\
              & C12 & 21000 & 2018 Sep 30 & 10 \\
\hline
 HST/WFC3 & H1 & ICDM20030 (F275W) & 2013 Oct 17 & 2.36 \\
                & H2 & ICDM20040 (F336W) & 2013 Oct 17 &  1.12\\
                & H3 & ICDM20050 (F555W) & 2013 Oct 17 & 0.96 \\
                & H4 & ID9609020 (F555W) & 2016 Oct 04 & 0.71 \\
                & H5 & ID9609010 (F814W) & 2016 Oct 04 & 0.78 \\
                & H6 & IDI102020 (F555W) & 2017 Dec 04 & 0.71 \\
                & H7 & IDI102010 (F814W) & 2017 Dec 04 & 0.78\\
                & H8 & IEB314020 (F555W) & 2021 Aug 19 & 0.71 \\
                & H9 & IEB314010 (F814W) & 2021 Aug 19 & 0.78 \\
                & H10 & IEB349020 (F438W) & 2021 Feb 15 &  0.71\\
\hline
        
 \label{Obs_log}
\end{tabular}
\end{table*}

\subsection{X-ray observations}

We analysed the {\it XMM-Newton} data using the standard software tools of {\sc sas}\footnote{https://www.cosmos.esa.int/web/xmm-newton/download-and-install-sas} (\textit{XMM-Newton} Science Analysis System) v21.0 \citep{2004ASPC..314..759G} and the latest Current Calibration Files (CCF) files were used for re-calibration. To create the calibrated event files, the {\sc epchain} and {\sc emchain} tasks were used for the EPIC PN and MOS cameras, respectively. After all the event files were obtained in the 0.3$-$10 keV energy range, the source and background counts were extracted from circular regions of 30$\arcsec$ and 60$\arcsec$ in radius using the {\sc evselect} task, respectively. The background regions were selected on source-free regions. The light curves of the source were also constructed using the {\sc etimeget} procedure in {\sc sas}. In XM1 and XM2 observations (see Table \ref{Obs_log}), almost half of the X-1 counts fall on the chip gap in the EPIC PN frames. Therefore, we used only MOS data when analysing these two observations. For XM3, we used both PN and MOS data.

The data reduction of the {\it Chandra} observations was performed using the {\it Chandra} Interactive Analysis of Observations ({\sc ciao}) v4.12 software package\footnote{https://cxc.cfa.harvard.edu/ciao/} and the {\it Chandra} Calibration Database (CALDB) v4.9.3 calibration files \citep{2006SPIE.6270E..1VF}. X-1 was located on the ACIS S3 (back-illuminated) chip. Using the {\sc chandra$\_$repro} script, the event files were created in the soft, hard, and total energy bands: 0.3$-$1.5 keV - soft (S), 1.5$-$10 keV - hard (H), 0.3$-$10 keV - total (T). At all energy ranges, the source and background counts were obtained using the {\sc specextract} task from circular regions of 5$\arcsec$ and 10$\arcsec$ in radius, respectively. The background photons were extracted from a location with no source contamination. The light curves were constructed using the {\sc dmextract} task in {\sc ciao} in the total energy range. 

The source was observed by the {\it Swift}/X-ray Telescope (XRT) 54 times between 2001 and 2023. The energy spectra of the source were extracted with the {\sc xselect} tool (v2.4) in the {\sc heasoft} package \citep{1995ASPC...77..367B}. While extracting the spectra, 35$\arcsec$ circular regions were selected both for the source and background. The spectral files were obtained in the 0.3$-$10 keV energy range. Due to insufficient statistics, only the net count rate of X-1 was calculated in the {\it Swift} data while examining the long-term variability of the source.

\subsection{HST observations and astrometry}

We used the HST/WFC3/UVIS observations to investigate the optical properties of the possible optical counterpart(s) of X-1. To search for optical counterpart candidates of the ULX in HST images, the relative astrometry between the {\it Chandra} and HST images was improved. The HST/WFC3 F336W (H2) drizzled image and C11 data were used to calculate the coordinate difference. Two reference sources that were detected by both telescopes were chosen while determining the relative shifts between two images. Then the shifts in R.A. and Dec. values were applied in order to calculate the corrected position of the ULX on the HST image. The corrected position of the source was estimated as R.A. = 01:36:51.0766, Dec. = +15:45:46.968 in the HST/WFC3 F336W drizzled image with an uncertainty of 0.3\arcsec.   

We performed point spread function (PSF) photometry using the {\scshape dolphot} v2.0 software. The data files used in this study were taken from the HST data archive.\footnote{https://mast.stsci.edu/search/ui/$\#$/hst} Stellar photometry was done by following the steps in the {\scshape dolphot} manual \citep{2000PASP..112.1383D, 2016ascl.soft08013D} using the HST/WFC3 images. We used the {\scshape wfc3mask} and {\scshape calcsky} tasks so that bad pixels were masked in the images and a sky image was created. Then the {\scshape dolphot} task was run to detect the possible optical sources and to perform photometry. The default {\scshape Dolphot} zero points and aperture corrections were adopted during the calculations. 

Determining the optical counterparts of ULXs allowed us to constrain the age and mass of the donor star. If it is assumed that the optical emission is dominated by the donor star, it is possible to determine the age of the donor star using theoretical isochrones. To estimate the mass of the donor stars, we obtained the colour-magnitude diagrams (CMDs) of the possible optical candidates of X-1. Using the Padova and Trieste Stellar Evolution Code ({\sc parsec})\footnote{http://stev.oapd.inaf.it/cgi-bin/cmd} isochrones for the HST/WFC3 wide filters, the CMD (F555W-F814W versus F555W) was created for the possible optical counterpart(s) and field stars located within 7$\arcsec$ around of X-1 (Fig. \ref{F:cmd}). While obtaining the CMD, we adopted the metallicity at solar value and the distance modulus as 29.9 mag. The PADOVA isochrones were corrected for the reddening of E(B-V)=0.15 mag.

\section{Results and discussion}
\label{results}

For this study, we investigated the X-ray spectral and temporal properties of X-1 in NGC 628 in detail using archival X-ray observations. NGC 628 has archival observations by {\it XMM-Newton}, {\it Chandra}, and {\it Swift} spanning 22 years. While many of these observations were used for the first time to study X-1, some of the data (XM1, XM2, C1, and C2 in Table \ref{Obs_log}) were re-analysed. Additionally, we searched the optical counterpart for X-1 using the archival HST/ WFC3 data and determined two optical counterpart candidates within the estimated error radius.

\subsection{X-ray spectral analysis}
\label{X-spec}

The spectral analyses of {\it XMM-Newton} and {\it Chandra} data were carried out in the 0.3$-$10 keV energy band using {\sc xspec} v12.13.1 \citep{1996ASPC..101...17A}. Prior to fitting, all spectra of X-1 were grouped to have a minimum of 20 counts per energy bin using the {\sc grppha} task. The spectra were then fitted with the power law ({\sc po}) and multi-colour disc black-body ({\sc diskbb}) models by multiplying with the absorption model ({\sc tbabs}) twice. While one of the absorption components was fixed to the Galactic value \cite[$0.05 \times 10^{22} \mathrm{cm}^{-2}$,][]{1990ARA&A..28..215D}, the other was left free to derive the intrinsic absorptions. The flux values were calculated using the {\sc cflux} convolution model in the 0.3$-$10 keV energy range. The best-fit model parameters are given in Table \ref{T:xpar}. 

\begin{table*}
        \centering
    \caption{Best-fit X-ray spectral parameters of X-1 in all epochs calculated with {\scshape tbabs*tbabs*po} and {\scshape tbabs*tbabs*diskbb} models.  }
        \begin{tabular}{ccccccccc}
        \hline \hline
 Obs.  &  Model & $N_{\mathrm{H}}$ $^{a}$  & $\Gamma$ / $kT_{\mathrm{in}}$ (keV) & N $^{b}$ & $\chi_{\nu}^{2}$ (dof) & $L_{\mathrm{X}}$ $^{c}$ & \multicolumn{2}{c} {$\alpha$$_{ox}$ $^{d}$} \\
        &    & ($10^{22}$ cm$^{-2}$) & & (10$^{-4}$) &  & (10$^{39}$ erg cm$^{-2}$) & c1 & c2 \\

\hline
\vspace{0.1cm}                                                                                                                                                                                  
C1      &{\sc po} &     0.00 &  $2.02_{-0.17}^{+0.18}$   &      $0.09_{-0.01}^{+0.01}$ &1.96 (13)      &        $0.63_{-0.06}^{+0.06}$ &       $-0.60_{-0.02}^{+0.02}$ &       $-0.55_{-0.02}^{+0.02}$ \\
\vspace{0.3cm}                                                                                                                                                                                                                                                                                                  
        &{\sc diskbb}   & 0.00  &       0.48 &  370 &   3.95 (13)       &       0.36    &...    &       ...\\
\vspace{0.1cm}                                                                                                                                                                                                                                                                                                  
C2      & {\sc po}      &       0.00 &  $1.76_{-0.11}^{+0.12}$  &       $0.24_{-0.01}^{+0.01}$  & 1.78 (38)       &       $1.90_{-0.10}^{+0.10}$  &       $-0.41_{-0.01}^{+0.01}$ &       $-0.36_{-0.01}^{+0.01}$ \\
\vspace{0.3cm}                                                                                                                                                                                                                                                                                                  
        &{\sc diskbb}   & 0.00  &       0.60 &  378 & 3.55 (38) &       1.06    &...    &       ...\\
\vspace{0.1cm}                                                                                                                                                                                                                                                                                                  
XM1     &{\sc po}       &       0.00 &  $1.94_{-0.14}^{+0.15}$  &       $0.21_{-0.01}^{+0.01}$  & 1.22 (93)       &       $1.54_{-0.14}^{+0.14}$  &       $-0.44_{-0.02}^{+0.02}$ &       $-0.39_{-0.01}^{+0.01}$ \\
\vspace{0.3cm}                                                                                                                                                                                                                                                                                                  
        &{\sc diskbb}   & 0.00  &       $0.60_{-0.01}^{+0.01}$  &       $75_{-5}^{+5}$  & 1.73 (93)       &       $0.86_{-0.06}^{+0.06}$  &       ...&... \\
\vspace{0.1cm}                                                                                                                                                                                                                                                                                          
XM2     &{\sc po}       & $0.03_{-0.02}^{+0.02}$ & $2.06_{-0.18}^{+0.22}$ &       $0.15_{-0.02}^{+0.02}$  &       1.25 (21) &     $0.99_{-0.12}^{+0.12}$  &       $-0.53_{-0.03}^{+0.03}$ &       $-0.47_{-0.02}^{+0.02}$ \\
\vspace{0.3cm}                                                                                                                                                                                                                                                                                                  
        &{\sc diskbb}   &       0.00 &  0.75    &       72      & 2.00 (21)     &       0.59    &  ...&   ...\\
\vspace{0.1cm}                                                                                                                                                                                                                                                                                                  
C5      &{\sc po}& $0.29_{-0.06}^{+0.07}$ &     $1.85_{-0.11}^{+0.13}$  & $0.73_{-0.07}^{+0.07}$ &         0.96 (12)      & $6.37_{-0.06}^{+0.06}$ & $-0.21_{-0.01}^{+0.01}$       &       $-0.16_{-0.01}^{+0.01}$ \\
\vspace{0.3cm}                                                                                                                                                                                                                                          
        &{\sc diskbb}   & $0.08_{-0.06}^{+0.07}$        &       $1.36_{-0.04}^{+0.04}$  &       $38_{-4}^{+4}$  & 0.75 (12)       &       $3.96_{-0.04}^{+0.04}$  &       ...&... \\
\vspace{0.1cm}                                                                                                                                                                                                                                                                                                  
C6      &{\sc po}& $0.07_{-0.02}^{+0.02}$ &     $1.59_{-0.05}^{+0.06}$ & $0.61_{-0.02}^{+0.02}$ &        1.14 (55)       & $5.23_{-0.03}^{+0.03}$ &       $-0.25_{-0.01}^{+0.01}$ & $-0.20_{-0.01}^{+0.01}$ \\
\vspace{0.3cm}                                                                                                                                                                                                                                                                                          
        &{\sc diskbb}   &       0.00 &  $1.43_{-0.02}^{+0.02}$  &       $28_{-1}^{+1}$  & 1.27 (55)       &       $3.68_{-0.02}^{+0.02}$  &...    &...            \\
\vspace{0.1cm}                                                                                                                                                                                                                                                                                                  
C7      &{\sc po}&      $0.23_{-0.05}^{+0.05}$ & $1.92_{-0.09}^{+0.10}$ & $0.97_{-0.07}^{+0.07}$ &        1.07 (21)       & $6.26_{-0.04}^{+0.04}$ &       $-0.21_{-0.01}^{+0.01}$ & $-0.16_{-0.01}^{+0.01}$       \\
\vspace{0.3cm}                                                                                                                                                                                                                                                                                                  
        &{\sc diskbb}&  $0.01_{-0.01}^{+0.01}$  &       $1.34_{-0.03}^{+0.03}$  &       $57_{-4}^{+4}$  &       0.99 (21)    &       $3.86_{-        0.03}^{+0.03}$  &       ...&... \\
\vspace{0.1cm}                                                                                                                                                                                                                                                                                          
C8      &{\sc po}&      $0.25_{-0.04}^{+0.05}$  & $1.99_{-0.09}^{+0.10}$ &       $0.95_{-0.07}^{+0.07}$  & 1.18 (22) &   $6.49_{-0.03}^{+0.03}$  & $-0.21_{-0.01}^{+0.01}$ & $-0.16_{-0.01}^{+0.01}$       \\
\vspace{0.3cm}                                                                                                                                                                                                                                                                                                  
        &{\sc diskbb}&  $0.01_{-0.01}^{+0.01}$  &       $1.35_{-0.03}^{+0.03}$  &       $44_{-3}^{+3}$  & 1.38 (22)       &       $3.89_{-0.03}^{+0.03}$  &       ...&...         \\
\vspace{0.1cm}                                                                                                                                                                                                                                                                                                  
C9  &{\sc po}   & $0.03_{-0.03}^{+0.07}$ &      $1.52_{-0.11}^{+0.13}$  & $0.56_{-0.06}^{+0.06}$ & 1.55 (11)      & $5.05_{-0.05}^{+0.05}$ &      $-0.26_{-0.01}^{+0.01}$ & $-0.21_{-0.01}^{+0.01}$ \\
\vspace{0.3cm}                                                                                                                                                                                                                                                                                          
        &{\sc diskbb}&  0.00    &       $1.43_{-0.04}^{+0.04}$  &       $32_{-3}^{+3}$  & 1.84 (11)       &       $3.58_{-0.05}^{+0.04}$  &       ...&... \\
\vspace{0.1cm}                                                                                                                                                                                                                                                                                                  
C10     &{\sc po}&      $0.21_{-0.03}^{+0.03}$ & $1.85_{-0.06}^{+0.06}$ & $0.73_{-0.04}^{+0.04}$ &         1.03 (52) & $5.57_{-0.02}^{+0.02}$ & $-0.24_{-0.00}^{+0.00}$   & $-0.19_{-0.00}^{+       0.00}$  \\
\vspace{0.3cm}                                                                                                                                                                                                                                                                                                  
        &{\sc diskbb}&  0.00&   $1.49_{-0.02}^{+0.02}$  &       $30_{-2}^{+2}$  & 0.77 (52)       &       $3.71_{-0.02}^{+0.02}$  &       ...&    ...\\
\vspace{0.1cm}                                                                                                                                                                                                                                                                                          
C11     &{\sc po}&      $0.05_{-0.02}^{+0.02}$ & $1.82_{-0.07}^{+0.08}$ & $0.44_{-0.02}^{+0.02}$  &       0.83 (35)       & $2.99_{-0.02}^{+0.02}$ & $-0.34_{-0.01}^{+0.01}$ & $-0.29_{-0.01}^{+0.01}$     \\
\vspace{0.3cm}                                                                                                                                                                                                                                                                                                  
        &{\sc diskbb}&  0.00    & 1.6   & 19 & 2.94 (35)        &       2.35    &...            &       ...                     \\
\vspace{0.1cm}                                                                                                                                                                                                                                                                                          
C12     &{\sc po}&      $0.03_{-0.03}^{+0.16}$ & $1.71_{-0.12}^{+0.15}$ & $0.60_{-0.07}^{+0.07}$ & 0.44 (6) & $4.20_{-0.05}^{+0.05}$      & $-0.26_{-0.03}^{+0.03}$ &       $-0.21_{-0.02}^{+0.02}$ \\
\vspace{0.3cm}                                                                                                                                                                                                                                                                                                  
        &{\sc diskbb}&  0.00    &$ 1.30_{-0.05}^{+0.04}$ &      $30_{-4}^{+4}$  & 1.21 (6) &      $2.85_{-0.06}^{+0.05}$  &               ...&...                 \\
\vspace{0.1cm}                                                                                                                                                                                                                                                                                                  
XM3     &{\sc po} &     $0.12_{-0.01}^{+0.01}$ & $2.14_{-0.04}^{+0.04}$ & $0.06_{-0.01}^{+0.01}$ &        0.92 (486) &    $3.24_{-0.01}^{+0.01}$  & $-0.34_{-0.00}^{+0.00}$ &       $-0.29_{-0.00}^{+0.00}$ \\
\vspace{0.3cm}                                                                                                                                                                                                                                                                                                  
        &{\sc diskbb} & 0.00    &       $1.04_{-0.01}^{+0.01}$  &       $79_{-2}^{+2}$  & 1.37 (486)      &       $2.07_{-0.01}^{+0.01}$  &               ...&... \\

\hline
        \end{tabular}
        \label{T:xpar}
        \\ Notes. The error values of the individual parameters were estimated at the 90\% confidence level. (a) Intrinsic absorption values towards the source. (b) Normalisation parameters of the {\scshape po} model in units of photons per kiloelectron volt per square centimetre per second at 1 keV and normalisation parameters of the {\scshape dıskbb} model, $N_{\mathrm{Disk}}$=[(r$_{\mathrm{in}}$ km$^{-1}$)/($D/10$ kpc)]$^{2} \times \cos i$. (c) The luminosity values were calculated in the 0.3$-$10 keV energy range. (d) Optical spectral index parameters in terms of the X-ray-UV correlation.
 \end{table*}

\begin{figure}
\begin{center}
\includegraphics[angle=0,scale=0.45]{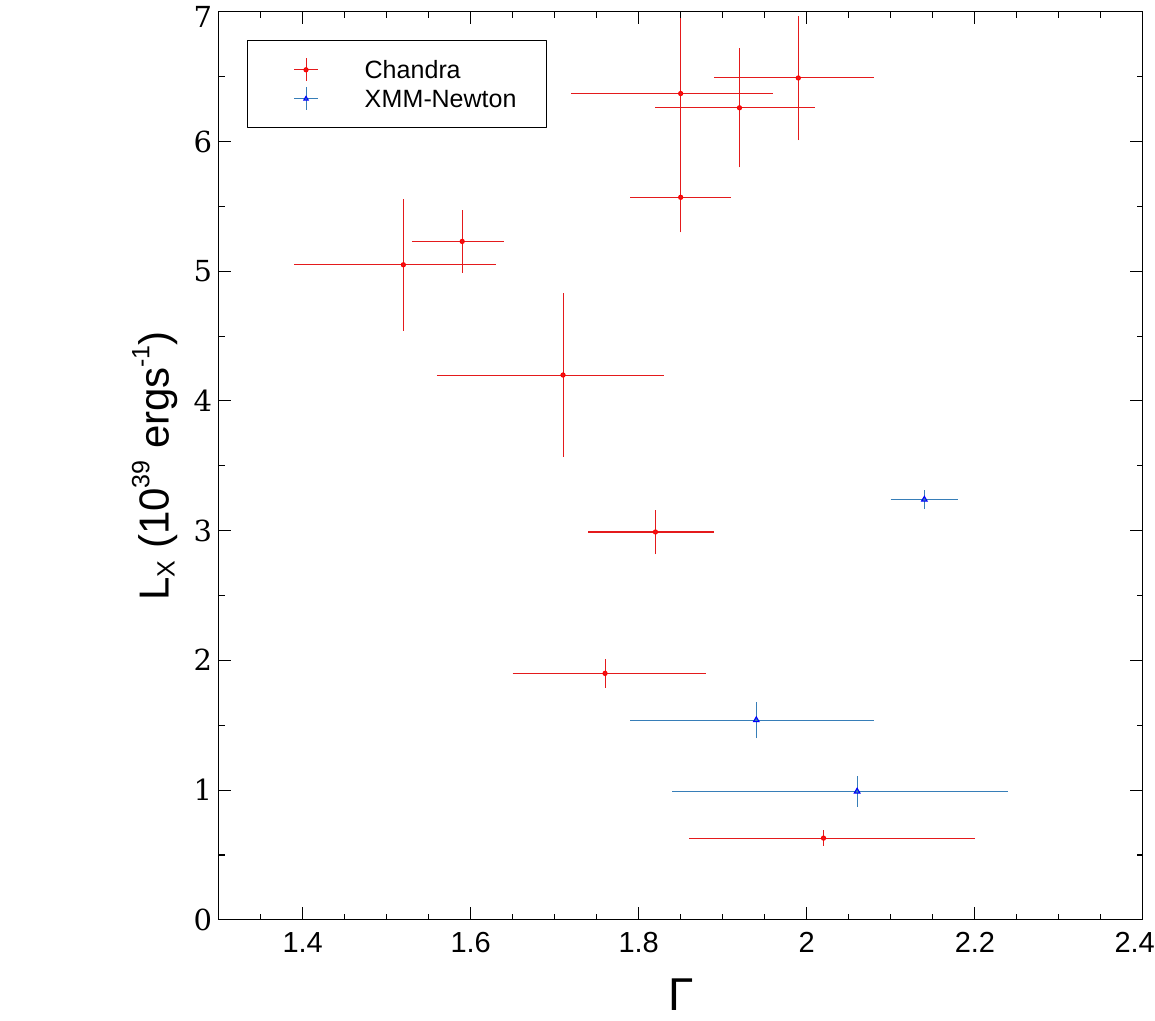}
\caption{Best-fit $\Gamma$ versus luminosity diagram of X-1. The $\Gamma$ values are derived from the {\sc po} model, and the luminosities were calculated in the 0.3$-$10 keV energy band.  }
\label{F:glx}
\end{center}
\end{figure}

The spectra of the source are generally well modelled with the {\sc po} model. The best-fit photon index ($\Gamma$) values slightly changed over the years, ranging from 1.52 to 2.14. Nonetheless, the luminosity of the source significantly changed by a factor of approximately ten. This trend can also be seen from the obtained $\Gamma - L_{\mathrm{X}}$ plot (Fig. \ref{F:glx}), where $\Gamma$ values are the spectral index derived from the {\sc po} model and $L_{\mathrm{X}}$ values are the unabsorbed luminosities in the 0.3$-$10 keV energy range. We performed a Pearson regression test to check whether any correlation is present between the two quantities (see Fig. \ref{F:glx}) using the {\sc scipy} package \citep{2020SciPy-NMeth}. The test did not yield any significant correlation (with a p value $\sim 0.3$).

We also fitted the spectra of X-1 with the {\sc po+diskbb} model to determine the spectral regime, whether soft ultraluminous, hard ultraluminous, or broadened disc \cite[three empirical classification for ULXs by][]{2013MNRAS.435.1758S}, in each epoch. Due to the low statistical quality, the parameters were poorly constrained except for the high-quality XM3 data. The best-fit model parameters derived the using XM3 data are given in Table \ref{T:po+diskbb}. Considering the best-fit parameters in Table \ref{T:po+diskbb} and by following the criteria given by \cite{2013MNRAS.435.1758S}, the spectral state of NGC 628 X-1 in XM3 data can be classified as hard ultraluminous.

\begin{table}
        \centering
    \caption{Best-fit parameters obtained from the {\sc tbabs*tbabs*(po+diskbb)} model for the spectrum of X-1 in XM3 data.}
        \begin{tabular}{lc}
        \hline \hline
 Parameter & Best-fit value \\
 \hline
\vspace{0.1cm}  
$N_{\mathrm{H}}$ ($10^{22}$ cm$^{-2}$) $^{a}$ &  $0.10_{-0.01}^{+0.01}$ \\
\vspace{0.1cm}  
$\Gamma$ & $1.90_{-0.05}^{+0.05}$ \\
\vspace{0.1cm}  
$kT_{\mathrm{in}}$ (keV) & $0.41_{-0.02}^{+0.01}$ \\
\vspace{0.1cm}  
$N_{\mathrm{po}}$ ($10^{-5}$) & $3.93_{-0.15}^{+0.15}$\\
\vspace{0.1cm}  
$N_{\mathrm{disk}}$ ($10^{-2}$) & $6.12_{-0.01}^{+0.01}$\\
\vspace{0.1cm}  
$\chi_{\nu}^{2}$ (dof) & 0.92 (484) \\
\vspace{0.1cm}  
$L_{\mathrm{X}}$ ($10^{39}$ erg s$^{-1}$) $^{b}$ & $2.98_{-0.09}^{+0.09}$ \\
\hline 
        \end{tabular}
        \label{T:po+diskbb}
        \\ Notes. The error values of the individual parameters were estimated at the 90\% confidence level. (a) Intrinsic absorption value towards the source. (b) The luminosity value was calculated in the 0.3$-$10 keV energy range. 
 \end{table}

\begin{figure}
\begin{center}
\includegraphics[angle=0,scale=0.35]{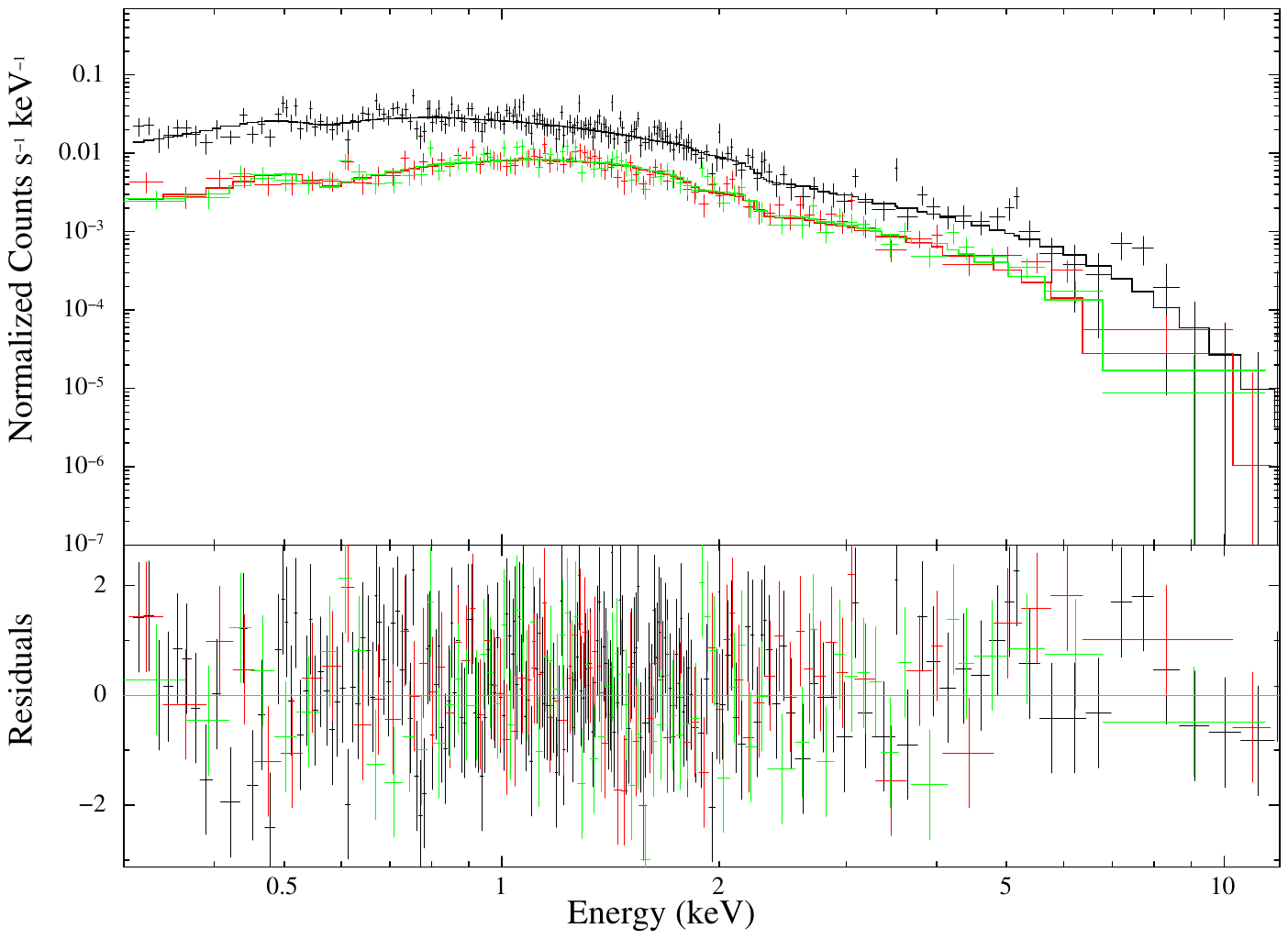}
\caption{{\it XMM-Newton} PN (black), MOS1(red), and MOS2 (green) energy spectrum of X-1 obtained using XM3 data. The spectrum was fitted with the {\scshape tbabs*tbabs*diskpbb} model.}
\label{F:spec}
\end{center}
\end{figure}

Usually, the spectra of ULXs are fitted with an extended disc black-body model ({\sc diskpbb} in {\sc xspec}) \citep{2005ApJ...631.1062K} in order to determine if the accretion disc is a standard disc \citep{1973A&A....24..337S} or a slim disc \citep{1988ApJ...332..646A,2001ApJ...549L..77W}. In the {\sc diskpbb} model (also known as the p-free model), the relation between disc temperature and radius is given by $T(R) \propto R^{p}$, where for a standard disc with sub-Eddington accretion, $p\sim0.75$. On the other hand, when the accretion rate increases near or above the super-Eddington regime, the disc becomes an advection-dominated slim disc, and the $p$ value reaches $\sim 0.5$ \citep{2000PASJ...52..133W,2007MNRAS.377.1187P}. We fitted the XM3 data that has the highest statistical quality with the {\sc diskpbb} model. The best-fit model parameters are given in Table \ref{T:diskpbb}, and the obtained energy spectrum of X-1 with the best-fit model is shown in Fig. \ref{F:spec}. The yielded $p$ value ($0.5$) indicates that the accretion disc in X-1 exhibits slim disc properties. Also, the inner disc temperature value calculated from the model ($kT_{\mathrm{in}} \sim 2.4$ keV) is very high, as seen in some ULXs \citep{2006PASJ...58..915V,2009MNRAS.397.1836G,2012PASJ...64..119I,2016MNRAS.459..455P,2020ApJ...890..166P}. The high temperature values are consistent with the slim disc model since the spectral hardening factor, $\kappa$, is considered to be higher in the case of a slim disc compared to a standard disc \citep{2003ApJ...596..421W}. 

 \begin{table}
        \centering
    \caption{Best-fit parameters obtained from the {\sc tbabs*tbabs*diskpbb} model for the spectrum of X-1 in XM3 data.  }
        \begin{tabular}{lc}
        \hline \hline
 Parameter & Best-fit value \\
 \hline
\vspace{0.1cm}  
$N_{\mathrm{H}}$ ($10^{22}$ cm$^{-2}$) $^{a}$ &  $0.01_{-0.01}^{+0.01}$ \\
\vspace{0.1cm}  
$kT_{\mathrm{in}}$ (keV) & $2.38_{-0.02}^{+0.02}$ \\
\vspace{0.1cm}  
$p$ & $0.50_{-0.01}^{+0.01}$ \\
\vspace{0.1cm}  
$N_{\mathrm{disk}}$ ($10^{-5}$) & $5.93_{-0.05}^{+0.05}$\\
\vspace{0.1cm}  
$\chi_{\nu}^{2}$ (dof) & 0.93 (485) \\
\vspace{0.1cm}  
$L_{\mathrm{X}}$ ($10^{39}$ erg s$^{-1}$) $^{b}$ & $2.91_{-0.09}^{+0.09}$ \\
\hline 
        \end{tabular}
        \label{T:diskpbb}
        \\ Notes. The error values of the individual parameters were estimated at the 90\% confidence level. (a) Intrinsic absorption value towards the source. (b) The luminosity value was calculated in the 0.3$-$10 keV energy range. 
 \end{table}

The mass of the compact object in X-1 can be estimated using the normalisation parameter calculated from the {\sc diskpbb} model. The apparent inner disc radius ($r_{\mathrm{in}}$) can be obtained by considering the normalisation parameter ($N$) with the following equation:

\begin{equation}
    r_{\mathrm {in}} (\mathrm{km}) = N^{1/2} (\cos i)^{-1/2} D_{\mathrm{10 kpc}} \,,
    \label{Eq1}
\end{equation}where $i$ is the inclination and $D$ is the distance in units of 10 kpc \citep{2000ApJ...535..632M}. The relation between the true ($R_{\mathrm{in}}$) and apparent inner disc radius ($r_{\mathrm{in}}$) is given by 

\begin{equation}
    R_{\mathrm {in}} (\mathrm{km}) = \xi \kappa^{2} r_{\mathrm{in}}\,,
    \label{Eq2}
\end{equation}
where the correction factor is $\xi = 0.353$ \citep{2008PASJ...60..653V} and the spectral hardening factor is $\kappa = 3$ \citep{2003ApJ...596..421W} for a slim disc.

For the spectrum of X-1 in XM3 data, the best-fit {\sc diskpbb} normalisation parameter was calculated as $N = (5.93\pm0.05)\times10^{-5}$. Assuming a moderate inclination of $i = 60^{\circ}$, the true inner disc radius for X-1 is estimated as $\sim 35$ km via Eq. (\ref{Eq1}) and Eq. (\ref{Eq2}). We then used the relation between the mass and inner disc radius \citep{2000ApJ...535..632M} by multiplying it with the minimum mass correction factor (1.2) for a slim disc \citep{2008PASJ...60..653V}:

\begin{equation}
    M = 1.2 (\frac{R_{\mathrm {in}} }{8.86 \alpha }) M_{\odot}\,,
\end{equation}where $\alpha=1$ for a non-rotating Schwarzschild BH and $\alpha = 1/6$ for an extremely rotating Kerr BH. Therefore, we derived the mass of the compact object in X-1 as $(5-28) M_{\odot}$ for both cases.

\subsection{X-ray temporal analysis}

The source exhibits quasi-periodic oscillation (QPO), which was detected previously in XM1, C1, and C2 data \cite[0.1$-$0.4 mHz,][]{2005ApJ...621L..17L}. We used the Lomb-Scargle method \citep{1976Ap&SS..39..447L, 1982ApJ...263..835S} in order to search for this QPO feature in other observations and to examine the short-term variability in the {\it XMM-Newton} and {\it Chandra} observations. The PN, MOS1, and MOS2 light curves of the source were combined with the {\sc lcmath} tool in the {\sc heasoft} package, and the combined light curves were used throughout the timing analyses of all the {\it XMM-Newton} observations. The Lomb-Scargle periodograms of the source were computed over the whole frequency range for each background-subtracted data by using the {\scshape LombScargle} tool \citep{2012cidu.conf...47V, 2015ApJ...812...18V} within the {\scshape astropy} package \citep{2022ApJ...935..167A}. The $3\sigma$ false alarm probability levels were estimated by following the method of \cite{2008MNRAS.385.1279B} and the bootstrap approach. The number of samples was set to 1000 when using the bootstrap method.

\begin{figure*}
\begin{center}
\includegraphics[angle=0,scale=0.45]{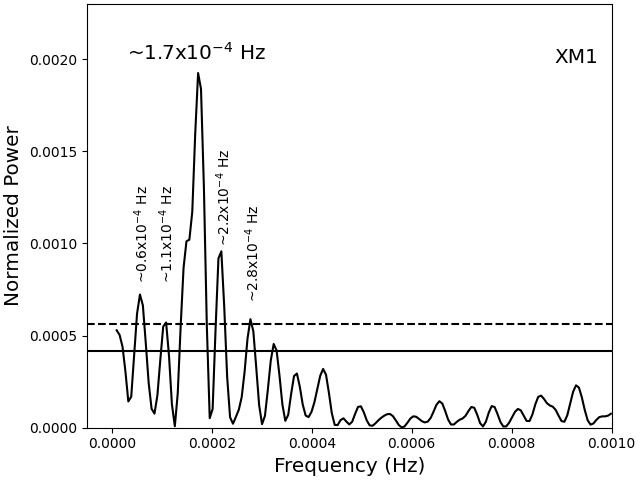}
\includegraphics[angle=0,scale=0.45]{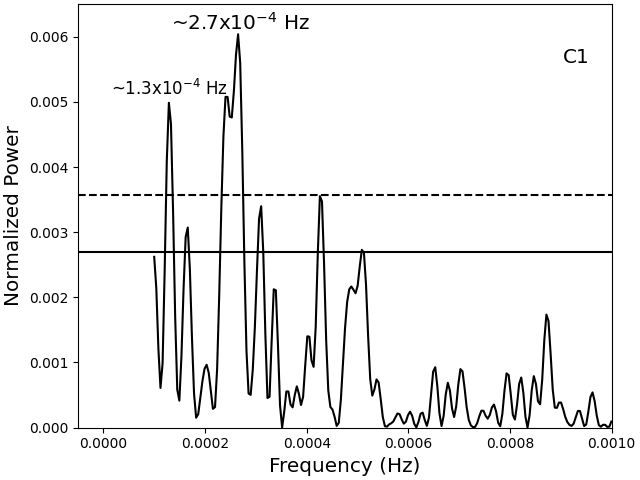}
\includegraphics[angle=0,scale=0.45]{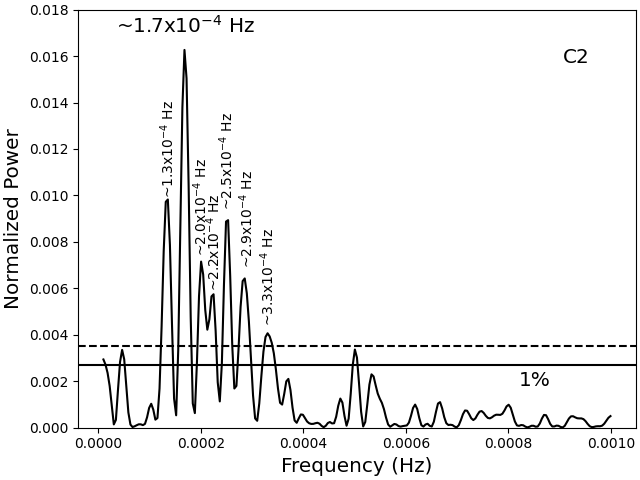}
\includegraphics[angle=0,scale=0.45]{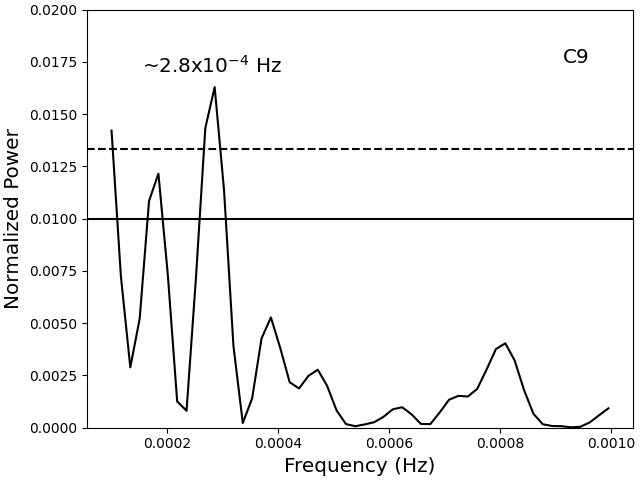}
\includegraphics[angle=0,scale=0.45]{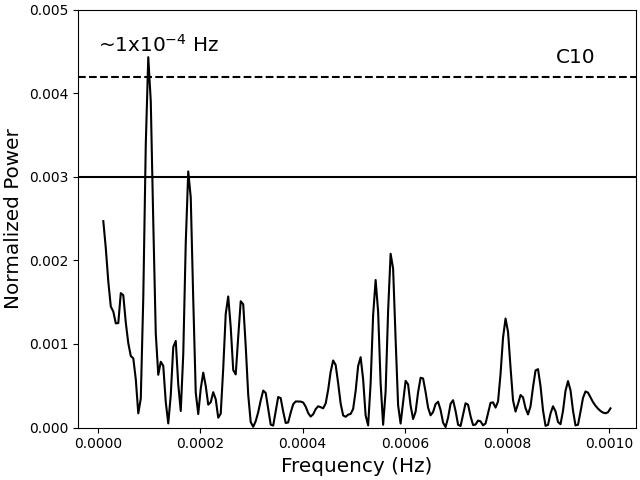}
\caption{Lomb-Scargle periodograms of X-1 obtained using XM1, C1, C2, C9, and C10 data. The black dashed and solid lines represent the $3\sigma$ false alarm levels estimated with the method of {\cite{2008MNRAS.385.1279B}} and the bootstrap approach, respectively. The centroid frequency values of the significant peaks are given in each panel.}
\label{F:LS}
\end{center}
\end{figure*}

The Lomb-Scargle periodograms of the source show significant QPO features ($>3\sigma$) in XM1, C1, C2, C9, and C10 data within the frequency range of $f = (1 - 3) \times 10^{-4} \mathrm{Hz}$. The obtained periodograms are shown in Fig. \ref{F:LS}. The solid and dashed lines in each panel represent the estimated $3\sigma$ false alarm probability levels computed with both methods. The centroid frequencies of the features are in good agreement with the previous results of \cite{2005ApJ...621L..17L}.

\begin{figure*}
\begin{center}
\includegraphics[angle=0,scale=0.45]{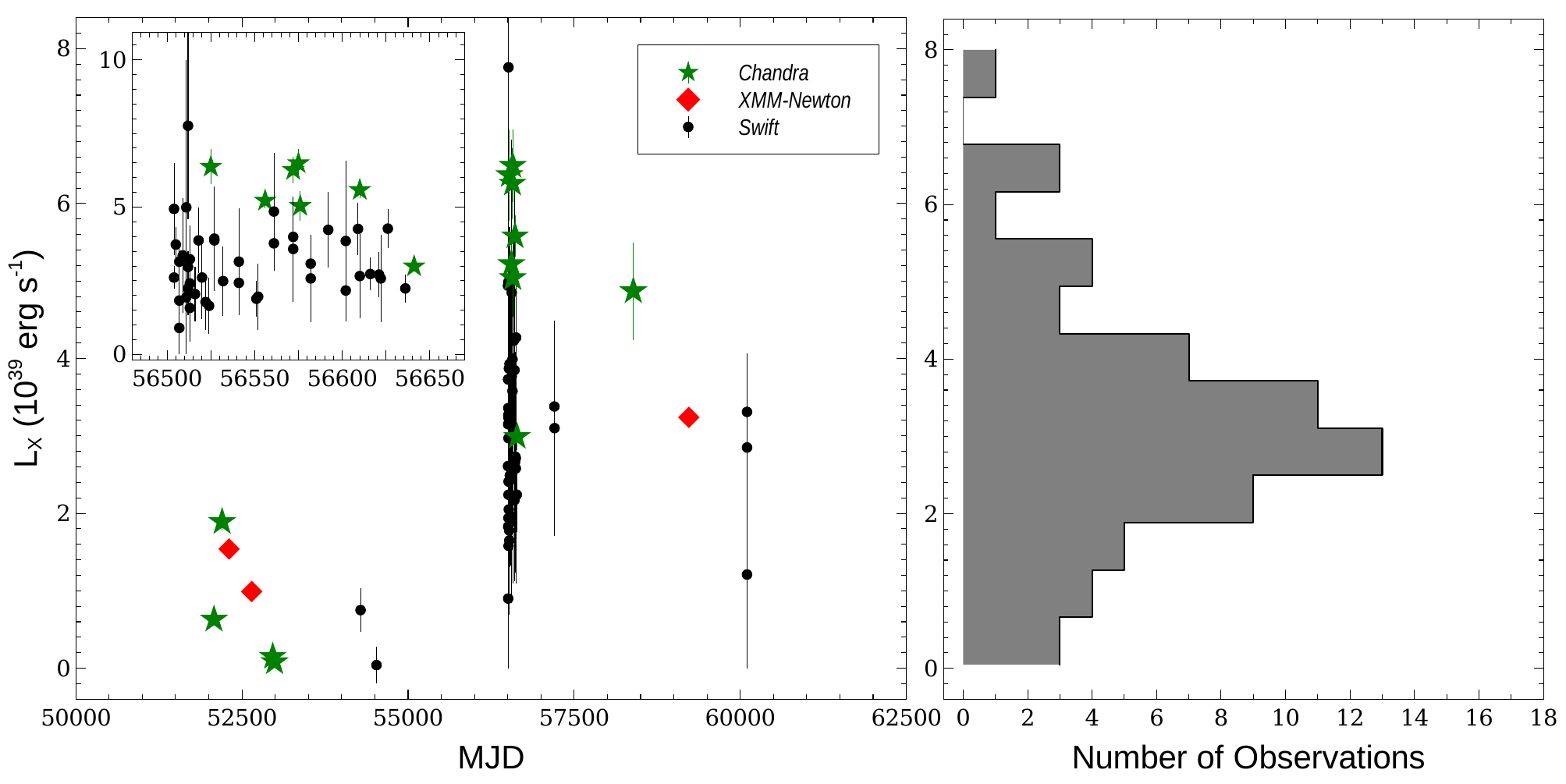}
\caption{Long-term light curve (left) and the histogram of the derived luminosities (right) of X-1 using the {\it Chandra}, {\it XMM-Newton}, and {\it Swift} data. The luminosity values were calculated in the 0.3$-$10.0 keV energy band. The inner panel in the light curve was given to separate the {\it Chandra} and {\it Swift} data for clarity.}
\label{F:lc}
\end{center}
\end{figure*}

Some ULXs are known to contain neutron stars via detection of pulsations \cite[PULXs;][]{2014Natur.514..202B, 2016ApJ...831L..14F, 2017MNRAS.466L..48I, 2018MNRAS.476L..45C, 2019MNRAS.488L..35S, 2020ApJ...895...60R, 2021MNRAS.503.5485Q}. The known PULXs generally exhibit a harder spectrum and significant long-term variability \citep{2017ApJ...836..113P, 2021A&A...649A.104G}. This long-term variability is sometimes in a structure of bimodal flux distribution, which may be explained by the onset of the  propeller regime \cite[e.g.][]{2016MNRAS.457.1101T}. Although we were not able to detect any similar pulsations, we compared the properties of the source with the properties of the well-known PULXs. The long-term light curve of X-1 and the histogram of the luminosities are given in Fig. \ref{F:lc}. Since the source has insufficient statistics to perform a spectral fitting in {\it Swift} and two {\it Chandra} (C3 and C4) observations, the count rates of the source in these observations were converted to unabsorbed flux values using the Portable, Interactive Multi-Mission Simulator ({\scshape pimms}).\footnote{https://heasarc.gsfc.nasa.gov/docs/software/tools/pimms.html} While performing the conversion, a power-law model was used by adopting the average $\Gamma$ value ($\sim$ 1.9) from Table \ref{T:xpar}, and the Galactic absorption was taken as $0.05\times10^{-22} \mathrm{cm}^{-2}$\citep{1990ARA&A..28..215D}. The light curve indicates that the source is variable, and the luminosity value of X-1 changed by a factor of $\sim$200 between the years 2001 and 2023.

\begin{figure}
\begin{center}
\includegraphics[angle=0,scale=0.50]{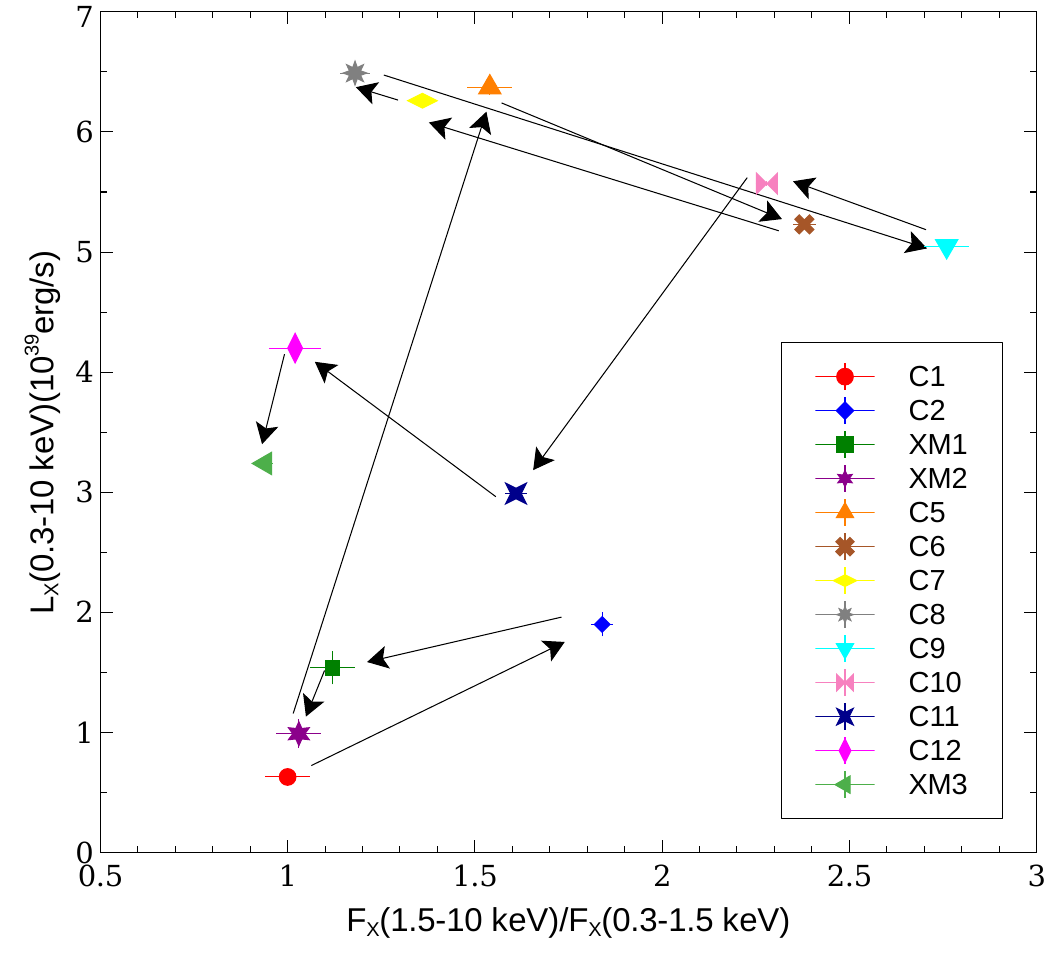}
\caption{Hardness-luminosity diagram of X-1. Temporal tracks are indicated by arrows. Each epoch is shown with different colours and symbols.}
\label{F:hld}
\end{center}
\end{figure}

In addition, we obtained a hardness-luminosity diagram (HLD) for X-1 (Fig. \ref{F:hld}), using all the available {\it XMM-Newton} and {\it Chandra} observations. While deriving the hardness ratio, the unabsorbed fluxes were calculated within the range of 0.3$-$1.5 keV and 1.5$-$10 keV for the soft and hard bands, respectively. The arrows in Fig. \ref{F:hld} indicate the transitions between the observations. In a recent study, HLDs of 17 ULXs were obtained and compared in order to constraint the nature of the compact source \citep{2021A&A...649A.104G}. They suggested that harder sources could harbour a strongly magnetised neutron star, whereas the softer ones could contain weakly magnetised neutron stars or BHs. Regarding the comparison of the X-1 HLD with that of the HLDs of ULXs in {\cite{2021A&A...649A.104G}}, we found that X-1 shows similarities with NGC 1313 X-2 (which exhibits softer spectra similar to NGC 628 X-1). Considering the variability and the similarity between the HLDs, the neutron star scenario for X-1 cannot be ruled out.

\subsection{Optical counterparts}

\begin{figure*}
\begin{center}
\includegraphics[angle=0,scale=0.42]{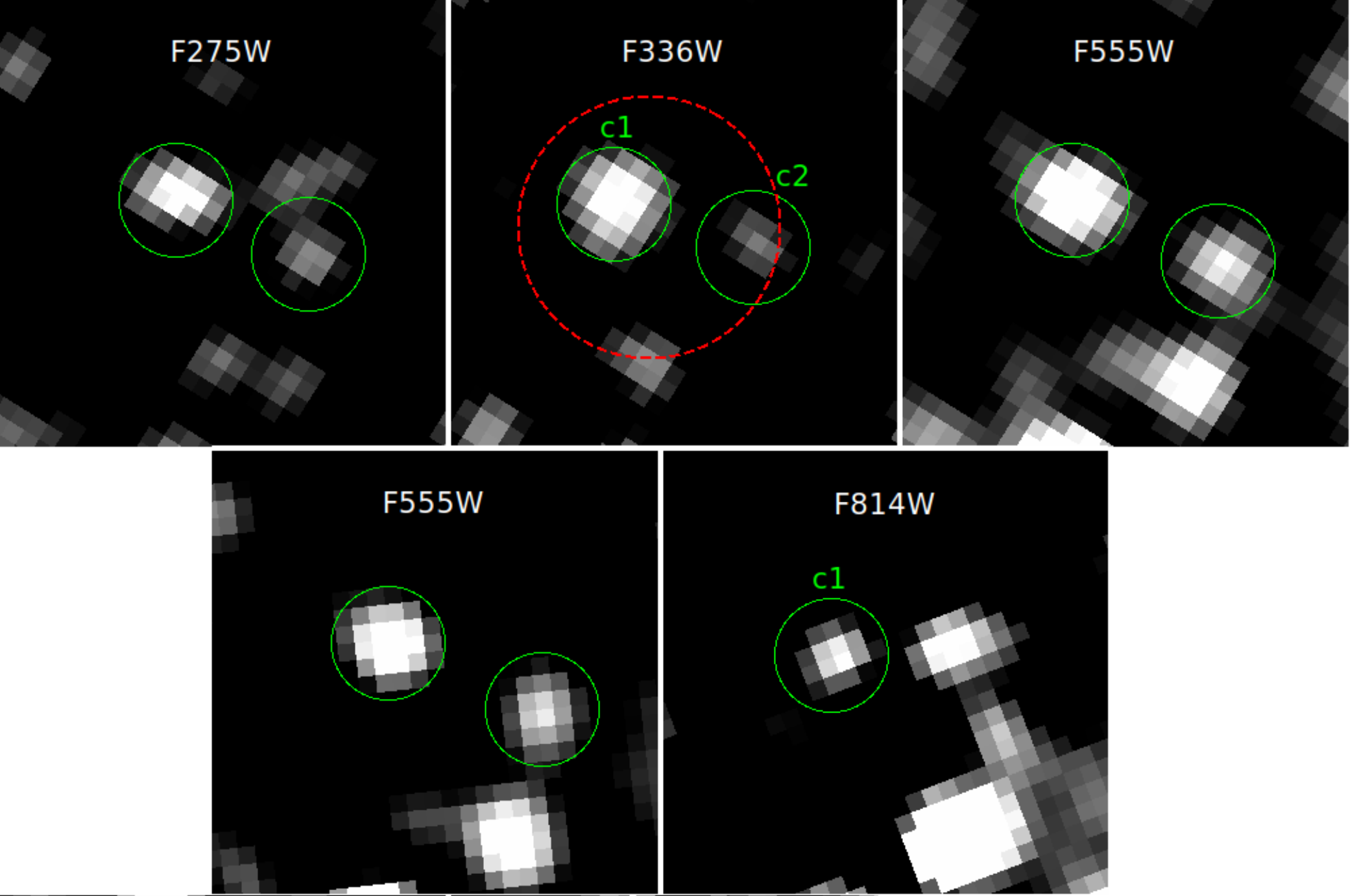}
\caption{Images from HST/WFC3 of the corrected positions of X-1 in three simultaneous filters (H1, H2 and H3, upper panel) and two simultaneous filters (H4 and H5, lower panel). The images have a size of $1\arcsec\times 1\arcsec$. The red dashed circle shows the astrometric error radii (0$\farcs$3). Green circles represent the positions of the optical counterpart candidates c1 and c2. North is up.}
\label{F:hst}
\end{center}
\end{figure*}

After the astrometric correction, two optical counterpart candidates (c1 and c2) were found within the error radius of 0$\farcs$3. The corrected position of X-1 and the counterpart candidates are shown in Fig. \ref{F:hst}. Although c1 was detected in all observations, c2 was not detected in the F814W filter images. The de-reddened HST/WFC3 magnitudes of c1 and c2 are given in Table \ref{T:opt_par}. In a previous study, the unique optical counterpart of X-1 was identified within the error radius after astrometric correction using the HST/ACS data \citep{2023MNRAS.519.4826A}. This counterpart coincides with source c1. Although the HST/WFC3 data are statistically better than previous observations, c1 and c2 sources were not detected in the WFC3/F438W filter (H3 observation in Table \ref{Obs_log}). Therefore, we adopted the previous ACS/F435W (B) magnitude of c1, which was calculated by \cite{2023MNRAS.519.4826A}, in the following calculations for c1. 

The absolute magnitudes of c1 and c2 were calculated as being in the range M$_{V}$=(-4.1)$-$(-3.6) and (-3.7)$-$(-3.1), respectively. Although the absolute magnitudes of both optical candidates are slightly dim, they are compatible with the values of previously studied ULXs (-8 < M$_{V}$ < -3, \cite{2017ARA&A..55..303K, 2017ASPC..510..395F}). Assuming that the donor star dominates the optical emission and using the absolute magnitudes and colour values, (V-I)$_{0}$, the spectral types of c1 were determined as being F$-$G type giant/supergiant \citep{1981Ap&SS..80..353S, 2001ApJ...558..309D}, and this is consistent with the previous study \citep{2023MNRAS.519.4826A}. The spectral type for c2 could not be calculated due to a lack of F438W (B) and F814W (I) magnitudes (which was also not detected in previous observations).

\begin{figure}
\begin{center}
\includegraphics[angle=0,scale=0.43]{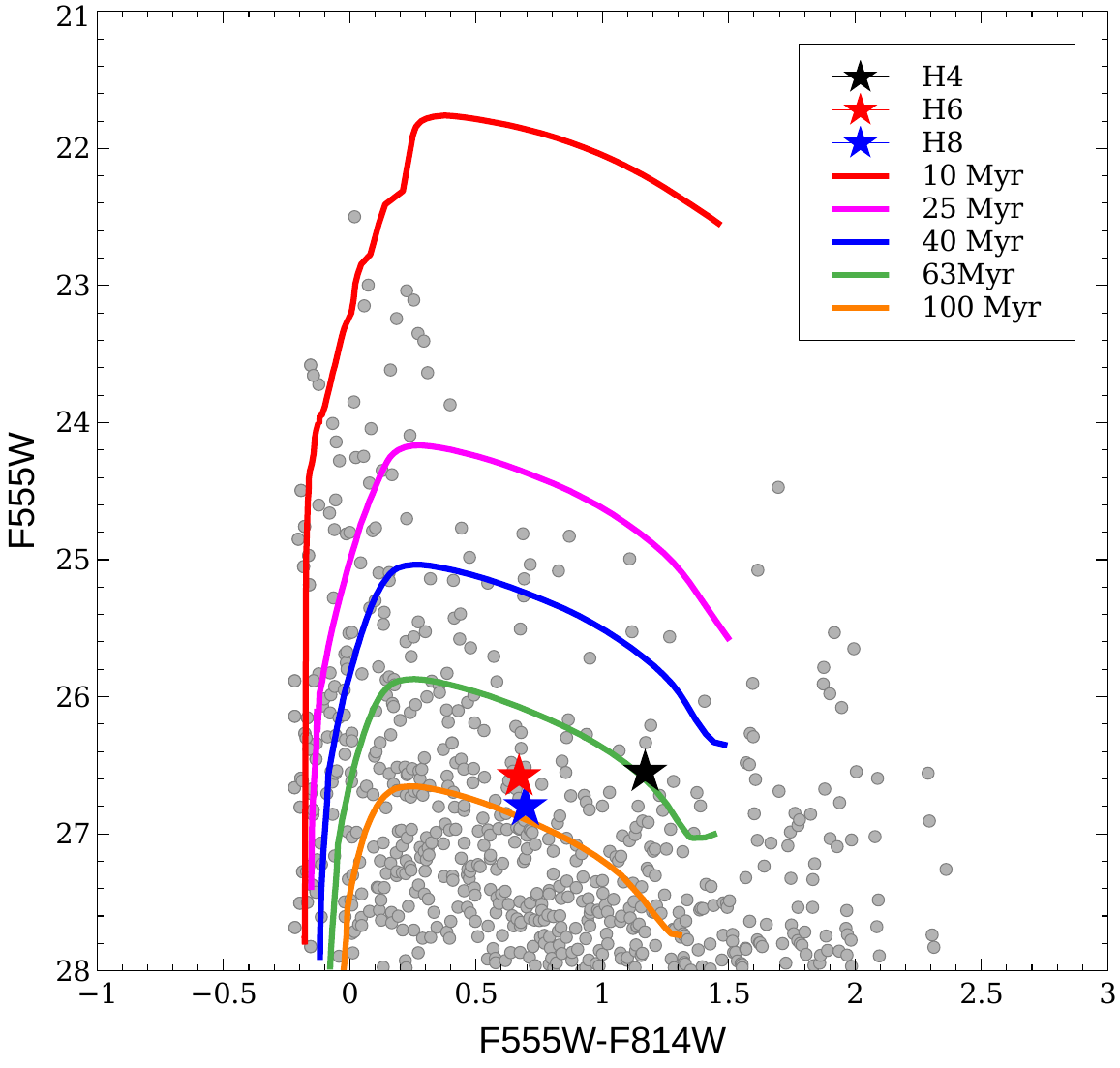}
\caption{Colour-magnitude diagram of X-1 from HST/WFC3. The optical counterpart candidate (c1) is shown as colourful stars in different observations, and field stars within 7$\arcsec$ around X-1 are shown as grey circles. The black, red, and blue colours represent magnitudes of c1 calculated using H4, H6, and H8 data, respectively. The PADOVA isochrones with different ages have been corrected for extinction of A$_{V}$=0.46 mag.}
\label{F:cmd}
\end{center}
\end{figure}

X-1 is located in a crowded region, but no star clusters associated with the source have been identified \citep{2023MNRAS.519.4826A}. Considering the CMD given in Fig. \ref{F:cmd}, if the optical emission comes from the donor star, the age of the star in this system is determined as $\geq$ 60 Myr. It can be seen that the field stars are divided into bright-young stars and older-fainter stars. The ages of the young and bright stars are 10$-$25 Myr, and their (V-I) colour values are between -0.2 and 0.5. The colour values of the older ones are between 0.1 and 2.0, and the colours of c1 are within this range as well.

It is known that the optical emission from ULXs can be dominated by a companion star and/or the accretion disc. There are a few helpful derivations that have been used to constrain the nature of the optical emission. The X-ray--to--optical flux ratio ($\xi$) can be used as a distinguishing parameter between low-mass X-ray binaries (LMXBs) and high-mass X-ray binaries (HMXBs). It is defined by \cite{1995xrbi.nasa...58V} as $\xi$=B$_{0}$+2.5$\times$log$F_{\mathrm{X}}$, where B$_{0}$ is the de-reddened B magnitude and F$_{X}$ is the X-ray flux in the 2$-$10 keV energy range in units of microjansky. Most ULXs have very high $\xi$ values ($\xi$ $\geq$20), indicating that the origin of optical emission from ULXs is similar to LMXBs rather than HMXBs \citep{ 2011ApJ...737...81T, 2019ApJ...875...68A}. This high value shows that the optical emission comes from the irradiated accretion disc. Since c1 does not have a WFC3/F438W magnitude, we calculated the $\xi$ value as $\sim$21 by adopting the ACS/F435W magnitude \cite[m$_{F435W}$=25.6,][]{2023MNRAS.519.4826A}. Due to a lack of simultaneous optical and X-ray observations, the flux value obtained from the XM2 observation (closest to the ACS/F435W observation) was used. In addition, by considering the high X-ray flux variability of X-1, the $\xi$ was calculated as 20.4 and 21.9 using the lowest derived flux (XM1) and the highest derived flux (C8) in the 2$-$10 keV energy band, respectively. In any case, the origin the optical emission of c1 is similar to LMXBs rather than HMXBs. 

Another X-ray--to--optical flux ratio is defined as log($F_{\mathrm{X}}$/$F_{\mathrm{opt}}$)= log$F_{\mathrm{X}}$+m$_{V}$/2.5+5.37 \citep{1982ApJ...253..504M}, where F$_{X}$ is the unabsorbed X-ray flux value in the 0.3$-$3.5 energy band, m$_{V}$ is the extinction-corrected visual magnitude. This ratio allows one to distinguish the active galactic nuclei (AGNs), BL Lac object, and cluster of galaxies. For these objects, this ratio is determined as being between -1 and 1.7 \citep{1982ApJ...253..504M, 1991ApJS...76..813S}. Using C9 and H3 observations, the log($F_{\mathrm{X}}$/$F_{\mathrm{opt}}$) values of c1 and c2 were calculated as 3.2 and 3.4, respectively. These ratios are greater than the ratios of AGNs, BL Lac objects, and cluster of galaxies, and they are similar to the optical counterparts of known ULXs \citep{2011ApJ...733..118Y, 2016ApJ...828..105A, 2023MNRAS.521.5298A}.

We also calculated the optical spectral index $\alpha$$_{ox}$ for the optical counterpart candidates as described by \cite{2019ApJ...873L..12S} (and references therein). This parameter gives the correlation between the UV flux density at 2500$\AA$ and the X-ray flux density at the 2 keV energy band. In each data, the X-ray flux density values of the ULX were obtained with a power-law model. By coupling each X-ray data with the only available HST/F275W observation (H1 in Table \ref{tab:obs-log}), the $\alpha$$_{ox}$ values for c1 and c2 were calculated. The obtained $\alpha$$_{ox}$ values for c1 and c2 using each X-ray data are given in Cols. 7$-$8 in Table \ref{T:xpar}. The $\alpha$$_{ox}$ for c1 and c2 were determined as being in the range of [-0.60, -0.21] and [-0.55, -0.16], respectively. These values are compatible with those found for the ULXs examined by \cite{2019ApJ...873L..12S} (see Table 1 and Fig. 1 in their paper).

\begin{figure}
\begin{center}
\includegraphics[angle=0,scale=0.55]{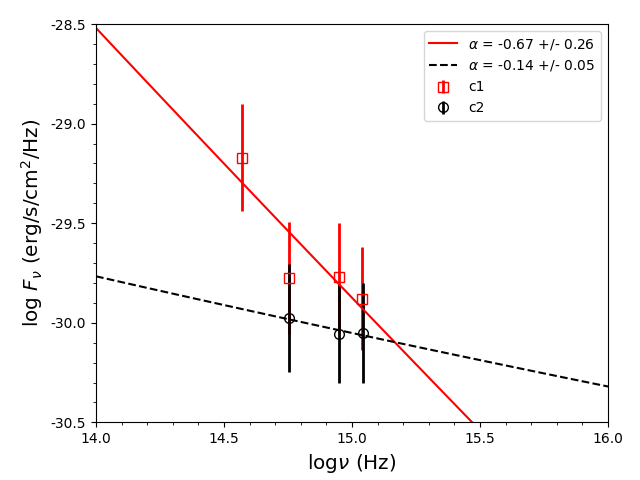}
\caption{Spectral energy distributions of the optical counterpart candidates for X-1. The c1 and c2 data are given as red squares and black circles, respectively. The SED models for c1 and c2 are shown as red solid and black dashed lines, respectively. The power-law indexes are $\alpha$=-0.67$\pm0.26$ for c1 and $\alpha$=-0.14$\pm0.05$ for c2.}
\label{F:sed}
\end{center}
\end{figure}

The spectral energy distributions (SEDs) were constructed using the HST observations with the F275W, F336W, F555W, and F814W filters for c1 and the F275W, F336W, and F555W filters for c2. The obtained SEDs are shown in Fig. \ref{F:sed}. The distributions of both counterpart candidates are consistent with a single power law (F$_{\nu} \propto \nu^{\alpha}$), similar to many ULXs \citep{2011ApJ...737...81T, 2013AstBu..68..139V, 2018ApJ...854..176V, 2019MNRAS.488.5935A}. The index value of the power-law models for the SEDs of c1 and c2 are $\alpha$=-0.67$\pm$0.26 and $\alpha$=-0.14$\pm$0.05, respectively. The SEDs with power-law index values between about -1 and 2 indicate that the optical emission is dominated by irradiation in the outer disc rather than the donor star.

\begin{table}
        \centering
        \caption{De-reddened magnitude values of the optical counterpart candidates of X-1 obtained with HST/WFC3 data.}
        \begin{tabular}{ccccc} 
                \hline  \hline
      
Obs.    &        \multicolumn{2}{c} {VEGAmag}   &        \multicolumn{2}{c} {M$_{V}$}                \\
        &       c1                      &       c2              &       c1 &       c2 \\
  \hline
H1      &       24.568  $\pm$   0.165   &       25.000  $\pm$   0.255   & ... & ... \\
H2      &       24.761  $\pm$   0.159   &       25.470  $\pm$   0.285   & ... & ...\\
H3      &       25.772  $\pm$   0.089   &       26.272  $\pm$   0.139   & -4.16 & -3.67\\
H4      &       26.090  $\pm$   0.117   &       26.480  $\pm$   0.150   & -3.85 & -3.45\\
H5      &       25.159  $\pm$   0.117   &       ...             &...&...\\
H6      &       26.122  $\pm$   0.121   &       26.455  $\pm$   0.155   & -3.81 & -3.48 \\
H7      &       25.690  $\pm$   0.206   &       ...             &...&...\\
H8      &       26.340  $\pm$   0.164   &       26.880  $\pm$   0.533  & -3.59 & -3.05 \\
H9      &       25.883  $\pm$   0.216   &       ...             &...&... \\
\hline 
    \end{tabular}
    \label{T:opt_par}
\end{table}

X-1 has four F555W observations between 2013 and 2021 (H3, H4, H6, and H8 data in Table \ref{T:opt_par}), and these allowed us to investigate the long-term magnitude variation of the counterpart candidates. The magnitude difference in the F555W (V) bands, V$_{\mathrm{min}}$ $-$ V$_{\mathrm{max}}$, for both candidates was calculated as 0.57$\pm$0.19 for c1 and 0.61$\pm$0.55 for c2. Similar high magnitude variations in the optical band that were seen in some ULXs were found to be correlated with the X-ray flux \citep{2012ApJ...750..152S,2020ApJ...893L..28V}. Unfortunately, we could not compare the optical variability of X-1 with its X-ray flux due to the lack of simultaneous (or quasi simultaneous) X-ray and optical data. The high variability in colour and absolute magnitude, $\xi$ value, and the high V$_{\mathrm{min}}$ $-$ V$_{\mathrm{max}}$ value of c1 indicate that the optical emission from this source is dominated by an irradiated accretion disc (\citep{2011ApJ...737...81T}; see Tables 5 and 6 in their paper).  Although c2 has a higher V$_{\mathrm{min}}$ $-$ V$_{\mathrm{max}}$, it is unreliable due to a high error value. Also, the fact that the source c2 is not detected in the F435W (B) and F814W (I) filters limits our knowledge about this source. However, it is difficult to ignore this source as an optical counterpart of X-1. Additional HST observations with different filters are required to investigate its optical properties in more detail.

\section{Summary}
\label{summary}

We have examined the X-ray properties of ULX X-1 in detail by using the available X-ray data of NGC 628. The previously reported {\it XMM-Newton} and {\it Chandra} observations were also re-analysed. To study the optical emission of X-1, we used the archival HST/WFC3 data. Below, we present a list that summarises the results.

\begin{itemize}

\item The X-ray spectra of X-1 are generally well fit with {\sc po}, with the spectral indexes in the range of $\Gamma$= 1.5$-$2.1. XM3 data, which provides the highest statistics. The spectra are also well fit with the {\sc po+diskbb} and {\sc diskpbb} models. For the former, the disc temperature is compatible with $kT_{\mathrm{in}} \approx 0.4$ keV, while for the latter it is $kT_{\mathrm{in}} \approx 2.4$ keV.

\item The previously detected QPO signal in X-1 was confirmed and detected in some of the newer datasets by using the Lomb-Scargle method. The frequency of the QPO signal is in the range of $f = (1-4) \times 10^{-4}$ Hz. 

\item The spectrum of X-1 yielded a good fit with the {\sc diskpbb} model, and the best-fit $p$ value ($0.5$) shows that the accretion disc exhibits slim disc properties. The mass of the BH in X-1 was derived as (5$-$28) $M_{\odot}$ by considering the {\sc diskpbb} model.
 
 \item The source exhibits a long-term variability by a factor of $\sim 200$ throughout the observations. The HLD of the source looks similar to the HLD of the previously studied PULX NGC 1313 X-2. This similarity makes it difficult to rule out the neutron star scenario for the nature of the compact object in X-1. The fact that the spectrum of the source is rather softer relative to other PULXs indicates that it may contain a weakly magnetised neutron star.   

 \item After astrometric correction, two optical counterpart candidates (c1 and c2) were identified for X-1. While c1 was detected in all observations but H10; c2 was not detected in the F814W and H10 observations.

 \item The c1 and c2 candidates are both brighter in the UV. The absolute magnitudes are M$_{V}$ $\sim$ -4 for c1 and M$_{V}$ $\sim$ -3 for c2. The spectral type of c1 was determined as being F-G giant/supergiant, taking into account its colour and absolute magnitude values. This is consistent with some known ULXs. Unfortunately, the spectral type for c2 could not be calculated due to the lack of the magnitude value in the adequate filter.

 \item The flux ratios (log$F_{\mathrm{X}}$/$F_{\mathrm{opt}}$) and optical spectral indexes, $\alpha$$_{ox}$, of the two counterpart candidates are consistent with ULXs rather than AGNs. Moreover, the $\xi$ values of c1 are consistent with LMXBs, indicating that the optical emission is dominated by the irradiated accretion disc.

 \item The derived power-law index for the SEDs of c1 and c2 are $\alpha$=-0.67 and $\alpha$=-0.14, respectively. These values are within the range given by \cite{2011ApJ...737...81T} and show that the optical emission is dominated by irradiation in the outer disc rather than the donor star.
\end{itemize}
   
Future X-ray observations of X-1 may reveal the true nature of the compact object in the system. In particular, observations carried out at higher energies will help us to study the X-ray emission of the source in more detail.

\begin{acknowledgements}
      This research was supported by the Scientific and Technological Research Council of Turkey (TÜBİTAK) through project number 122C042. We thank E. Sonbas and K. S. Dhuga for their useful comments.
\end{acknowledgements}

%
%
 \bibliographystyle{aa} 
 \bibliography{sample}

\end{document}